\documentclass[cameraready]{Interspeech}

\title{A Dynamic Knowledge Distillation Framework for Mitigating Spatial Ambiguity in Lightweight Dual-Channel Speech Enhancement}

\author[affiliation={1,2}]{Yifei}{Yang}
\author[affiliation={1,2}]{Zheng}{Wang}
\author[affiliation={3}]{Yu}{Sun}
\author[affiliation={1,2}]{Wentao}{Hua}
\author[affiliation={1,2}]{Xiaobin}{Rong}
\author[affiliation={1,2}]{Kai}{Chen}
\author[affiliation={1,2}]{Jing}{Lu}

\address{
    $^1$ Key Laboratory of Modern Acoustics, Nanjing University, China \\
    $^2$ NJU-Horizon Intelligent Audio Lab, Horizon Robotics, China \\
    $^3$ R\&D Centre, Samsung Electronics (China), China
}

\email{\{yifei.yang, zheng.wang\}@smail.nju.edu.cn, yuyu.sun@samsung.com, \{wentao.hua, xiaobin.rong\}@smail.nju.edu.cn, \{chenkai, lujing\}@nju.edu.cn}

\keywords{Speech enhancement, dual-channel, spatial ambiguity, knowledge distillation}

\usepackage{comment}
\usepackage{multirow}
\usepackage{subcaption}
\usepackage{threeparttable}
\usepackage{url}

\begin{document}

\maketitle

\begin{abstract}
Multi-channel speech enhancement based on deep learning has achieved significant progress. While keeping low complexity, lightweight dual-channel models exploit spatial cues to outperform single-channel models on edge devices. However, their performance often degrades for closely-spaced sources due to unconstrained spatial reliance. To mitigate this, we propose a Dynamic Spatial-aware Knowledge Distillation (D-SKD) framework that enhances spatial discrimination. Specifically, D-SKD introduces a dynamic arbitration mechanism into the distillation process with a spatially invariant single-channel teacher, enabling the dual-channel model to adaptively regulate spatial reliance without explicit angular supervision. Experiments on dual-channel GTCRN and other models show that D-SKD consistently improves performance for closely-spaced sources while preserving quality elsewhere. Requiring no additional inference overhead or auxiliary inputs, it is well suited for lightweight deployment.
\end{abstract}

\section{Introduction}

Speech enhancement (SE) aims to recover clean target speech in noisy environments. While single-channel SE relies on temporal or spectral characteristics to suppress interference \cite{Luo2019Conv, Le2021DPCRN}, multi-channel SE further exploits spatial information captured by microphone arrays, introducing an additional dimension to separate target speech from interferers \cite{Gannot2017Consolidated}. Due to this advantage, multi-channel SE has become an important front-end module for applications such as teleconferencing, smart glasses, and automatic speech recognition (ASR) systems \cite{Ren2021Causal, Xu2024Fovnet, boeddeker2018front}.

Recent advances in deep neural networks (DNNs) have significantly improved multi-channel SE by enabling data-driven modeling of complex nonlinear relationships in adverse acoustic environments \cite{Wang2018Supervised, Li2025Advances}. In particular, end-to-end multi-channel architectures can jointly model spatial and spectro-temporal information without relying on hand-crafted signal processing modules, and they consistently outperform single-channel counterparts in most conditions \cite{Tesch2022Role, haeb2025microphone, Quan2024SpatialNet}. Among these approaches, lightweight dual-channel models are especially attractive for low-footprint edge devices, as they offer a favorable balance between enhancement performance and computational cost \cite{wang2025lightweight}.

However, the performance gains of multi-channel models are not consistent across all scenarios. In closely-spaced source conditions, where the target speaker and the interferer are located in similar directions, spatial cues become less discriminative. In such cases, unconstrained reliance on spatial information can introduce \textit{spatial ambiguity}, leading to severe performance degradation and even making multi-channel models underperform their single-channel counterparts \cite{chen2019multi, gu2020enhancing, gu2020multi}. This issue is particularly critical for lightweight models, as the limited capacity of these architectures makes them more vulnerable to unreliable spatial cues. 

Existing studies have attempted to address this problem by explicitly or implicitly exploiting the source distance to regulate the use of spatial features. A representative approach is to jointly estimate the target speech and the source distance in a multi-task framework, and then switch between multi-channel and single-channel processing modes accordingly \cite{chen2019multi}. However, such model integration schemes may introduce additional memory and deployment overhead. Another line of work uses direction-of-arrival (DOA) information to guide attention mechanisms over spatial features across varying source distances \cite{Gu2019Neural}, but this requires prior localization information, which may not always be available in practical applications. To the best of our knowledge, existing methods still rely on auxiliary geometric information or additional model components to mitigate spatial ambiguity.

\begin{figure*}[t]
  \centering
  \includegraphics[width=\textwidth]{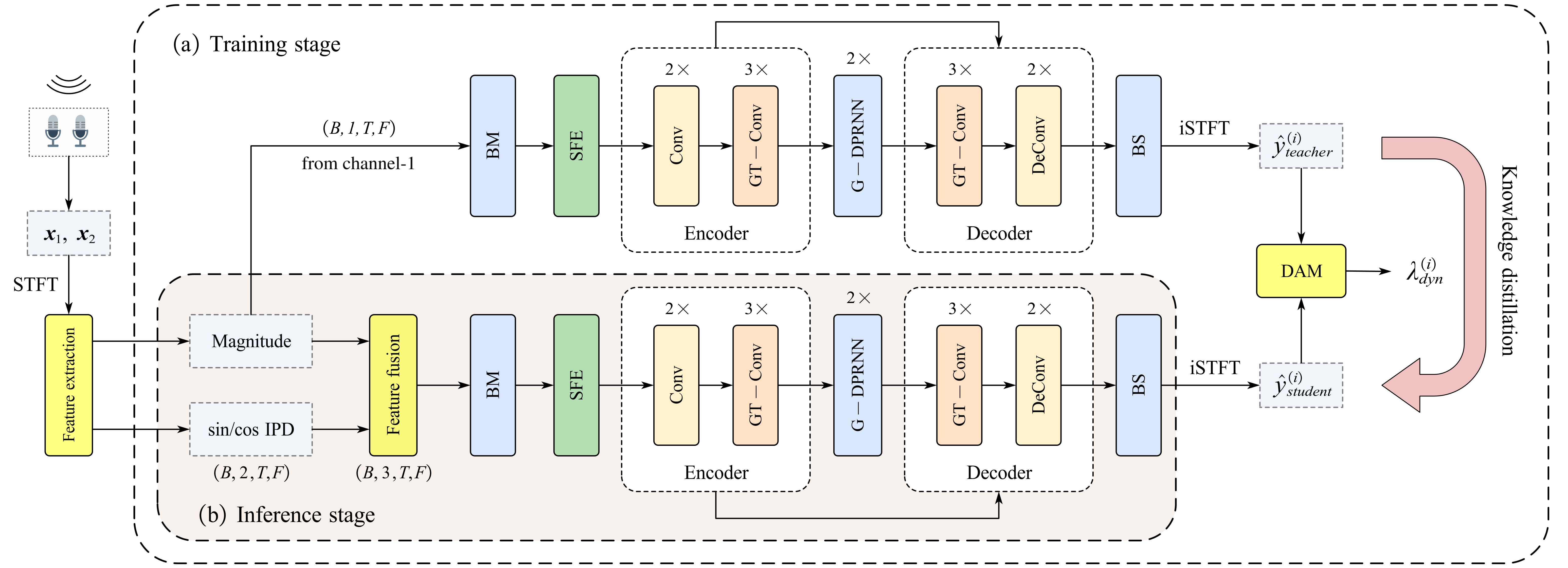} 
  \caption{Overall architecture of D-SKD. (a) Training stage, including the complete teacher and student branches, where their outputs guide the distillation process via DAM. (b) Inference stage, where only the student branch is retained.}
  \label{fig:1}
\end{figure*}

To address this limitation, we propose a Dynamic Spatial-aware Knowledge Distillation (D-SKD) framework for lightweight dual-channel SE. The key idea is to use the spatially invariant single-channel model as the teacher to provide a stable reference when spatial cues are unreliable, while allowing the dual-channel student model to retain the advantage of spatial information when it is informative. Specifically, D-SKD introduces a dynamic arbitration mechanism into the distillation process, enabling the student to adaptively regulate its reliance on spatial cues during training without explicit source-angle supervision. During inference, no additional parameters or auxiliary information are required. We apply the framework to the dual-channel variant of GTCRN \cite{Rong2024GTCRN} (DC-GTCRN), an effective end-to-end lightweight SE model. Experiments show that D-SKD consistently improves DC-GTCRN performance in closely-spaced source scenarios ($0^\circ$--$15^\circ$) while largely preserving performance in other conditions. Additional experiments on other state-of-the-art (SOTA) lightweight models further demonstrate the generalizability of D-SKD.

\section{Problem formulation}

To facilitate the investigation of the spatial ambiguity problem, we consider a reverberant scenario with one target speaker and one interferer. The time-domain signal captured by a 2-element microphone array is formulated as
\begin{equation}
x_k(n) = h_k(n) \ast s(n) + v_k(n), \quad k=1,2,
\label{eq:1}
\end{equation}
where $\ast$ denotes linear convolution. For the $k$-th microphone at time $n$, $x_k(n)$ is the captured mixture, $s(n)$ is the clean speech, $h_k(n)$ is the room impulse response (RIR) from the target speaker, and $v_k(n)$ denotes the \textbf{reverberant interference}. Under the narrow-band approximation \cite{avargel2007system}, Eq.~(\ref{eq:1}) can be reformulated in the short-time Fourier transform (STFT) domain as
\begin{equation}
X_k(t,f) = H_k(f)S(t,f) + V_k(t,f),
\label{eq:2}
\end{equation}
where $t$ and $f$ are the time-frame and frequency-bin indices. Without loss of generality, \textbf{the first microphone} is selected as the reference channel.

\section{Proposed method}

As illustrated in Figure~\ref{fig:1}, the proposed D-SKD framework builds upon the GTCRN baseline to improve dual-channel performance in closely-spaced source scenarios without compromising overall effectiveness. Section~\ref{sec:3.1} details the GTCRN architecture and its input feature representations adopted in different conditions. Section~\ref{sec:3.2} introduces D-SKD in detail, including the Dynamic Arbitrator Module (DAM), loss function, and training strategy. Specifically, DAM evaluates sample difficulty on an instance-wise basis during training to generate the corresponding distillation weights. In addition to the primary task loss, a dynamic distillation loss is introduced to guide the network to adaptively regulate its reliance on spatial information. Notably, D-SKD operates exclusively during training and requires no additional computational overhead or model parameters during inference.

\subsection{Model overview}\label{sec:3.1}

Based on the U-Net architecture, GTCRN processes time-frequency signals through a sequence of modules: Band Merging (BM), Subband Feature Extraction (SFE), encoder, G-DPRNN bottleneck, decoder, and Band Splitting (BS). To reduce redundancy, BM compresses high-frequency subbands into equivalent rectangular bandwidth (ERB) subbands, which are reconstructed in BS. SFE integrates adjacent subbands into the channel dimension for efficient convolutional processing. The encoder maps these features to high-dimensional embeddings, which are passed to the G-DPRNN for time-frequency modeling, and finally decoded to estimate the complex ratio mask (CRM) \cite{williamson2015complex} for waveform reconstruction via iSTFT.

For the single-channel version (SC-GTCRN), the magnitude spectrum of the first microphone is selected as the input, containing the principal spectral patterns. For the dual-channel model (DC-GTCRN), in addition to the magnitude spectrum of the first microphone, the sine and cosine of inter-channel phase difference (sin/cos IPD) are incorporated. These two types of features are concatenated along the channel dimension to serve as the final input. Such a sin/cos IPD structure provides robust spatial cues for separating the target speech from interference, while avoiding the numerical discontinuity problem caused by high-frequency phase wrapping \cite{wang2018combining, gu2021complex}.

\subsection{D-SKD training framework}\label{sec:3.2}

Spatial ambiguity is a persistent challenge in multi-channel speech separation and enhancement \cite{liao2023improvement, kovalyov2023dsenet}. As the target speaker and the interferer spatially approach each other, the spatial diversity relied upon by multi-channel models progressively diminishes, leading to continuous performance degradation. Since spatial cues typically offer higher discriminability than spectral features in most scenarios, DNNs are prone to developing an over-reliance on spatial information, resulting in insufficient emphasis on spectral features. As a consequence, the model attempts to rely on ambiguous or invalid spatial information for separation, ultimately failing in closely-spaced source scenarios. Under these extreme conditions, the performance of the single-channel model---which relies solely on spectral information---can be regarded as a reliable reference. Inspired by knowledge distillation \cite{hinton2015distilling}, we introduce a distillation loss to guide the dual-channel student in learning spatially invariant patterns from the single-channel teacher, thereby encouraging stronger reliance on spectral features in closely-spaced source scenarios.

\subsubsection{Dynamic arbitrator module (DAM)}

Blindly distilling knowledge from a single-channel teacher to a dual-channel student risks degrading the inherent spatial processing capabilities of the student. To address this, we introduce DAM to activate the distillation process only when the teacher outperforms the student. Specifically, after forward propagation during training, the hybrid losses for both teacher and student are initially computed per sample in the batch:
\begin{align}
  \mathcal{L}_{T}^{(i)} &= \mathcal{L}_H(\hat{y}_{T}^{(i)}, y_{\mathrm{clean}}^{(i)}), \label{eq:3} \\
  \mathcal{L}_{S}^{(i)} &= \mathcal{L}_H(\hat{y}_{S}^{(i)}, y_{\mathrm{clean}}^{(i)}), \label{eq:4}
\end{align}
where $i$ is the sample index, $\hat{y}_{T}^{(i)}$ and $\hat{y}_{S}^{(i)}$ are the enhanced outputs of the teacher and student,  $y_{\mathrm{clean}}^{(i)}$ is the training target of the current sample, and $\mathcal{L}_H$ is the hybrid loss (Eq.\ref{eq:9} in Section~\ref{sec:3.2.2}). We then evaluate their relative performance gap:
\begin{equation}
  \Delta L^{(i)} = \frac{\mathcal{L}_{S}^{(i)} - \mathcal{L}_{T}^{(i)}}{\mathcal{L}_{T}^{(i)}}. \label{eq:5}
\end{equation}
Since the loss function is utilized to characterize model performance, a smaller value of $\Delta L^{(i)}$ indicates superior performance of the student model. Finally, the dynamic distillation weight is calculated:
\begin{equation}
  \lambda _{dyn}^{\left( i \right)} = \mathrm{ReLU}(\tanh(\gamma \cdot \Delta L^{(i)})),
  \label{eq:6}
\end{equation}
where $\tanh$ is applied to map the values to $(-1, 1)$, and $\mathrm{ReLU}$ subsequently generates a $[0, 1)$ gating value. The arbitration sensitivity coefficient, $\gamma$, controls the $\tanh$ slope to regulate how $\lambda _{dyn}^{\left( i \right)}$ responds to $\Delta L^{(i)}$. When $\Delta L^{(i)} \leq 0$ (student outperforms), the distillation weight becomes 0. Conversely, when $\Delta L^{(i)} > 0$ (teacher outperforms), the weight rapidly ascends toward 1.

\subsubsection{Loss function}\label{sec:3.2.2}

The student model's loss comprises a primary task loss and a distillation loss, both utilizing the hybrid loss function \cite{Rong2024GTCRN}:
\begin{align}
  \mathcal{L}_{\mathrm{task}}^{(i)} = \mathcal{L}_H(\hat{y}_{S}^{(i)}, y_{\mathrm{clean}}^{(i)}). \label{eq:7} \\
  \mathcal{L}_{\mathrm{dist}}^{(i)} = \mathcal{L}_H(\hat{y}_{S}^{(i)}, \hat{y}_{T}^{(i)}). \label{eq:8}
\end{align}
Here, $\mathcal{L}_H$ combines SI-SNR and complex spectral losses:
\begin{equation}
\begin{split}
  \mathcal{L}_H =&~ \alpha \mathcal{L}_{\mathrm{SISNR}}(\hat{s}, s) + (1-\beta) \mathcal{L}_{\mathrm{mag}}(\hat{S}, S) \\
  & + \beta (\mathcal{L}_{\mathrm{real}}(\hat{S}, S) + \mathcal{L}_{\mathrm{imag}}(\hat{S}, S)),
\end{split}
\label{eq:9}
\end{equation}
where $\hat{s}$ and $s$ denote the estimated and target waveforms, respectively, while $\hat{S}$ and $S$ denote their complex spectra. $\alpha$ and $\beta$ are empirically set to 0.01 and 0.3. Consequently, the total loss for a batch of size $B$ is defined as
\begin{equation}
  \mathcal{L}_{\mathrm{total}} = \frac{1}{B}\sum_{i=1}^{B}\mathcal{L}_{\mathrm{task}}^{(i)} + \mu \cdot \frac{1}{B}\sum_{i=1}^{B}\left[ \lambda _{dyn}^{\left( i \right)} \cdot \mathcal{L}_{\mathrm{dist}}^{(i)} \right],
  \label{eq:10}
\end{equation}
where $\mu$ is a tunable global distillation weight, and $\lambda _{dyn}^{\left( i \right)}$ is the dynamic distillation weight calculated by DAM.

\subsubsection{Training and inference}

We employ a pre-trained SC-GTCRN as the teacher and a DC-GTCRN with the same backbone configuration as the student. During training, independent forward propagations yield their respective losses (Eq.~\ref{eq:3}, Eq.~\ref{eq:4}). The DAM evaluates their relative performance to generate the dynamic distillation weight $\lambda _{dyn}^{\left( i \right)}$. Subsequently, the total loss (Eq.~\ref{eq:10}) updates only the student, keeping the teacher frozen. During inference, the teacher and the DAM are discarded, leaving only the dual-channel student model.

\section{Experiments and results}

\subsection{Dataset}

We simulate a 16~kHz dual-channel speech enhancement dataset using speech and noise from the DNS-3 corpus \cite{reddy2021interspeech}. The training set comprises 50,000 noisy-clean pairs, each with a duration of 10~s. RIRs are simulated using the image method \cite{allen1979image} for a two-microphone array with a spacing of 4~cm. Room dimensions ($L, W \in [3, 10]$~m, $H \in [2.5, 3]$~m) and RT60 (0.1--0.4~s) are randomly sampled. Sources are placed coplanar with the array at randomly selected distances of 0.5, 1, 2, or 3 m, maintaining $\geq$ 0.5~m wall clearance and a $0^\circ$--$180^\circ$ angular separation. Training SNR ranges from $-$5 to 5~dB. A $500$-sample validation set follows identical settings. Three test sets are constructed at SNR = \SI{5}{\decibel} for the subsequent experiments:

\begin{itemize}
\item \textbf{Segmented-angle:} Five DOA difference intervals ($0^\circ$--$15^\circ$, $15^\circ$--$30^\circ$, $30^\circ$--$60^\circ$, $60^\circ$--$90^\circ$, $90^\circ$--$180^\circ$) with $500$ samples uniformly distributed within each interval at a 1~m distance.
\item \textbf{Average-angle:} $1000$ samples with uniformly distributed DOA differences ($0^\circ$--$180^\circ$) at a 1~m distance.
\item \textbf{Angular-scan:} In a fixed acoustic environment selected from the average-angle test set, noise is fixed at ($0^\circ$, 1~m) while the target speaker is moved from $5^\circ$ to $175^\circ$ in $10^\circ$ increments, yielding 200 samples per position.
\end{itemize}

\subsection{Implementation details}

Dual-channel mixtures are processed via STFT (32 ms Hann window, 16 ms shift) to extract the magnitude spectrum from the first channel, alongside the sin/cos IPD, as input features. The default hidden size (h.size) for convolutional and G-DPRNN layers is 16, with 32 and 64 available for larger variants. 

Both the teacher and student models are trained using the Adam optimizer \cite{kingma2014adam}. The learning rate is initialized to $0.001$, featuring a linear warmup over the first $20$ epochs followed by a cosine annealing decay schedule. Training proceeds for $200$ epochs with a batch size of $8$, randomly sampling 10,000 items per epoch. The teacher is optimized via the hybrid loss (Eq.~\ref{eq:9}), while the student employs the D-SKD loss (Eq.~\ref{eq:10}), where $\mu=1$ and $\gamma$ is adjusted according to the scale of the network.

\subsection{Results and discussion}

To assess the enhancement performance, we use two objective metrics: Perceptual Evaluation of Speech Quality (PESQ) \cite{rix2001perceptual} and Short-Time Objective Intelligibility (STOI) (\%) \cite{taal2010short}.

\subsubsection{Ablation study}

\begin{table*}[t]
  \caption{Ablation study results across different angular intervals.}
  \label{tab:1}
  \centering
  \begin{minipage}{\linewidth}
    \setlength{\tabcolsep}{5pt} 
    \resizebox{\linewidth}{!}{
      \begin{tabular}{@{} c @{\hspace{2em}} c @{\hspace{2em}} cc @{\hspace{1em}} cc @{\hspace{1em}} cc @{\hspace{1em}} cc @{\hspace{1em}} cc @{\hspace{1em}} cc @{}}
        \toprule
        \multirow{2}{*}{IDs} & \multirow{2}{*}{Methods \& Setup} & 
        \multicolumn{2}{c@{\hspace{1em}}}{$0^\circ$--$15^\circ$} & 
        \multicolumn{2}{c@{\hspace{1em}}}{$15^\circ$--$30^\circ$} & 
        \multicolumn{2}{c@{\hspace{1em}}}{$30^\circ$--$60^\circ$} & 
        \multicolumn{2}{c@{\hspace{1em}}}{$60^\circ$--$90^\circ$} & 
        \multicolumn{2}{c@{\hspace{1em}}}{$90^\circ$--$180^\circ$} & 
        \multicolumn{2}{c}{Avg.} \\
        \cmidrule(lr){3-4} \cmidrule(lr){5-6} \cmidrule(lr){7-8} \cmidrule(lr){9-10} \cmidrule(lr){11-12} \cmidrule(lr){13-14}
        & & 
        PESQ & STOI (\%) & 
        PESQ & STOI (\%) & 
        PESQ & STOI (\%) & 
        PESQ & STOI (\%) & 
        PESQ & STOI (\%) & 
        PESQ & STOI (\%) \\
        \midrule
        -  & Noisy              & 1.285 & 66.28 & 1.278 & 65.95 & 1.285 & 66.35 & 1.268 & 65.70 & 1.281 & 65.76 & 1.273 & 65.83 \\
        -  & SC-GTCRN           & \textbf{1.858} & \textbf{74.58} & 1.855 & 74.26 & 1.870 & 74.92 & 1.847 & 74.27 & 1.857 & 74.28 & 1.855 & 74.41 \\
        -  & DC-GTCRN           & 1.742 & 71.80 & 1.985 & 75.85 & 2.219 & \textbf{80.22} & \textbf{2.371} & \textbf{82.40} & \textbf{2.499} & 83.79 & 2.314 & 81.18 \\
        \midrule
        1 & D-SKD (proposed)         & 1.793 & 72.96 & \textbf{1.999} & \textbf{76.14} & \textbf{2.223} & 80.01 & 2.358 & 82.13 & 2.498 & \textbf{83.95} & \textbf{2.316} & \textbf{81.28} \\
        2 & w/ HAM & 1.802 & 73.32 & 1.968 & 75.85 & 2.178 & 79.69 & 2.317 & 81.69 & 2.450 & 83.40 & 2.284 & 80.92 \\
        3 & w/o DAM      & 1.834 & 73.56 & 1.901 & 74.67 & 2.009 & 76.83 & 2.073 & 77.91 & 2.178 & 79.51 & 2.072 & 77.82 \\
        \bottomrule
      \end{tabular}
    }
    
    \vspace{1ex}
    {\raggedright \footnotesize
    $^*$ Different angular intervals correspond to the \textbf{segmented-angle} test sets. ``Avg.'' represents the \textbf{average-angle} test set. 
    
    $^*$ \textbf{Bold} values indicate the best score in each test set.\par}
  \end{minipage}
\end{table*}

    To validate the effectiveness of D-SKD, Table~\ref{tab:1} presents a comparison of different distillation strategies. In ID-1, it can be observed that the proposed D-SKD framework effectively enhances the performance of DC-GTCRN on the $0^\circ$--$15^\circ$ test set with negligible performance degradation observed across other angular intervals. Benefiting from the mitigation of spatial ambiguity, the performance on the average-angle test set exhibits a marginal improvement. In ID-2, DAM is replaced with a Hard Arbitrator Module (HAM), where $\lambda _{dyn}^{\left( i \right)}$ only takes values of 0 or 1, rather than being computed via Eq.~\ref{eq:6}. While it yields a marginal improvement over ID-1 on the $0^\circ$--$15^\circ$ test set, this localized gain comes at the cost of degraded performance in other intervals. In ID-3, the DAM is completely removed. Although the model effectively utilizes spectral information to achieve enhancement in closely-spaced source scenarios, the capability to exploit spatial information degrades significantly.

\begin{figure}[t]
  \centering
  \begin{subfigure}[b]{0.495\linewidth}
    \centering
    \includegraphics[width=\linewidth]{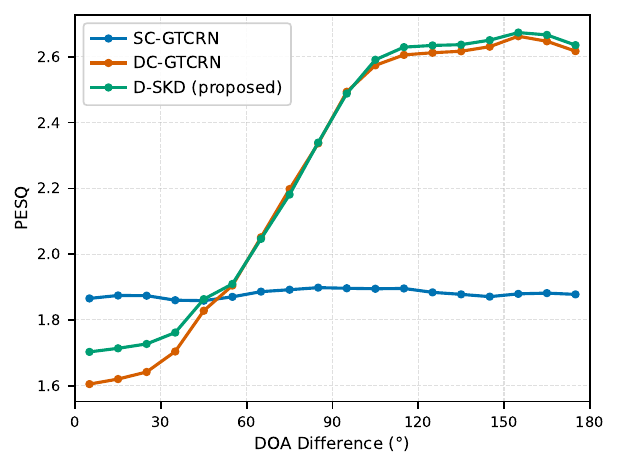}
    \caption{PESQ results}
    \label{fig:sub_a}
  \end{subfigure}
  \hfill
  \begin{subfigure}[b]{0.495\linewidth}
    \centering
    \includegraphics[width=\linewidth]{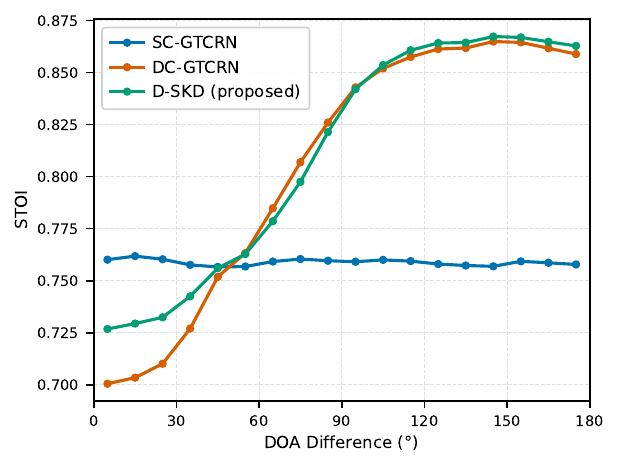}
    \caption{STOI results}
    \label{fig:sub_b}
  \end{subfigure}  
  \caption{Scanning results on the angular-scan test set.}
  \label{fig:2}
\end{figure}

Figure~\ref{fig:2} illustrates the evaluation results on the angular-scan test set. The performance degradation of DC-GTCRN as the sources spatially approach each other can be clearly observed. In contrast, D-SKD mitigates the spatial ambiguity problem, yielding a clear improvement in the performance of DC-GTCRN in closely-spaced source scenarios. This implies that DC-GTCRN successfully internalizes the spatially invariant knowledge from SC-GTCRN, learning to adaptively regulate its reliance on spatial information. Audio samples at a $5^\circ$ DOA difference are available\footnote{\url{https://muefy.github.io/D-SKD-Audio-Demo}}.

\subsubsection{Generalization on different network scales}

Table~\ref{tab:2} evaluates D-SKD across different scales of GTCRN. For each network scale, $\gamma$ is empirically tuned via preliminary validation. Larger models exhibit lower absolute losses and larger relative loss gaps in Eq.~\ref{eq:5}; thus, smaller $\gamma$ values are used to stabilize distillation. The results show that D-SKD consistently improves performance across all three evaluated network scales. Notably, D-SKD yields a highly significant improvement for DC-GTCRN with a hidden size of 32, elevating its $0^\circ$--$15^\circ$ performance to match SC-GTCRN while concurrently enhancing performance in other intervals. This indicates that the model learns more comprehensive spectral and spatial patterns through D-SKD and effectively balances the reliance on them, thereby further validating the immense potential of the proposed framework in mitigating the spatial ambiguity problem.

\subsubsection{Generalization on different lightweight models}

We further apply D-SKD to two SOTA lightweight models, LiSenNet \cite{yan2025lisennet} and UL-UNAS \cite{rong2026ul}. Both models adopt the same training configurations and input feature representations as GTCRN. Table~\ref{tab:3} presents the evaluation results on different test sets. D-SKD effectively mitigates spatial ambiguity across these lightweight models. It consistently improves performance on the $0^\circ$--$15^\circ$ test set while maintaining robust performance in other conditions without noticeable degradation.

\begin{table}[t]
  \caption{Results on different network scales.}
  \label{tab:2}
  \centering
  \setlength{\tabcolsep}{4pt} 
  \resizebox{\linewidth}{!}{
  \begin{tabular}{@{} c c c c cc cc cc @{}}
    \toprule
    \multirow{2}{*}[-0.5ex]{Setup} & 
    \multirow{2}{*}[-0.5ex]{\begin{tabular}{@{}c@{}}MACs \\ (M/s)\end{tabular}} & 
    \multirow{2}{*}[-0.5ex]{\begin{tabular}{@{}c@{}}Para. \\ (k)\end{tabular}} & 
    \multirow{2}{*}[-0.5ex]{D-SKD} & 
    \multicolumn{2}{c}{$< 15^\circ$} & 
    \multicolumn{2}{c}{$> 15^\circ$} & 
    \multicolumn{2}{c}{Avg.} \\
    \cmidrule(lr){5-6} \cmidrule(lr){7-8} \cmidrule(lr){9-10}
    & & & & 
    PESQ & STOI  & 
    PESQ & STOI  & 
    PESQ & STOI  \\
    \midrule    
    \multirow{3}{*}[-0.5ex]{\begin{tabular}{@{}c@{}}h.size = 16 \\[0.5ex] $\gamma$ = 1.0\end{tabular}} 
    & 28.46 & 23.19 & Teacher & 1.858 & 74.58 & 1.856 & 74.32 & 1.855 & 74.41 \\
    \cmidrule(lr){2-10}
    & \multirow{2}{*}{31.90} & \multirow{2}{*}{23.67} & $\times$ & 1.742 & 71.80 & \textbf{2.397} & \textbf{82.64} & 2.314 & 81.18 \\
    & & & \checkmark & \textbf{1.793} & \textbf{72.96} & 2.393 & 82.46 & \textbf{2.316} & \textbf{81.28} \\
    \midrule    
    \multirow{3}{*}[-0.5ex]{\begin{tabular}{@{}c@{}}h.size = 32 \\[0.5ex] $\gamma$ = 0.5\end{tabular}} 
    & 89.04 & 77.84 & Teacher & 1.989 & 76.39 & 1.993 & 76.21 & 1.992 & 76.27 \\
    \cmidrule(lr){2-10}
    & \multirow{2}{*}{94.37} & \multirow{2}{*}{78.81} & $\times$ & 1.925 & 75.60 & 2.621 & 84.62 & 2.532 & 83.43 \\
    & & & \checkmark & \textbf{1.987} & \textbf{76.35} & \textbf{2.632} & \textbf{84.77} & \textbf{2.554} & \textbf{83.74} \\
    \midrule    
    \multirow{3}{*}[-0.5ex]{\begin{tabular}{@{}c@{}}h.size = 64 \\[0.5ex] $\gamma$ = 0.1\end{tabular}} 
    & 313.70 & 281.62 & Teacher & 2.216 & 79.47 & 2.225 & 79.51 & 2.222 & 79.51 \\
    \cmidrule(lr){2-10}
    & \multirow{2}{*}{322.81} & \multirow{2}{*}{283.54} & $\times$ & 2.111 & 78.45 & 2.789 & \textbf{86.25} & 2.715 & 85.24 \\
    & & & \checkmark & \textbf{2.142} & \textbf{78.65} & \textbf{2.802} & 86.16 & \textbf{2.722} & \textbf{85.24} \\    
    \bottomrule
  \end{tabular}
  }
\end{table}

\begin{table}[t]
  \caption{Results on different lightweight models.}
  \label{tab:3}
  \centering
  
  \begin{minipage}{\linewidth}
    \setlength{\tabcolsep}{4pt} 
    \resizebox{\linewidth}{!}{
      \begin{tabular}{@{} c c c c cc cc cc @{}}
        \toprule
        \multirow{2}{*}[-0.5ex]{Models} & 
        \multirow{2}{*}[-0.5ex]{\begin{tabular}{@{}c@{}}MACs \\ (M/s)\end{tabular}} & 
        \multirow{2}{*}[-0.5ex]{\begin{tabular}{@{}c@{}}Para. \\ (k)\end{tabular}} & 
        \multirow{2}{*}[-0.5ex]{D-SKD} & 
        \multicolumn{2}{c}{$< 15^\circ$} & 
        \multicolumn{2}{c}{$> 15^\circ$} & 
        \multicolumn{2}{c}{Avg.} \\
        \cmidrule(lr){5-6} \cmidrule(lr){7-8} \cmidrule(lr){9-10}
        & & & & 
        PESQ & STOI  & 
        PESQ & STOI  & 
        PESQ & STOI  \\
        \midrule    
        \multirow{3}{*}[-0.5ex]{LiSenNet} 
        & 55.65 & 36.77 & Teacher & 1.960 & 75.86 & 1.954 & 75.87 & 1.959 & 75.87 \\
        \cmidrule(lr){2-10}
        & \multirow{2}{*}{55.77} & \multirow{2}{*}{36.78} & $\times$ & 1.841 & 73.46 & \textbf{2.598} & \textbf{83.91} & \textbf{2.509} & \textbf{82.61} \\
        & & & \checkmark & \textbf{1.885} & \textbf{74.23} & 2.592 & 83.83 & 2.506 & 82.51 \\
        \midrule    
        \multirow{3}{*}[-0.5ex]{UL-UNAS} 
        & 34.35 & 171.33 & Teacher & 2.080 & 77.66 & 2.074 & 77.49 & 2.077 & 77.55 \\
        \cmidrule(lr){2-10}
        & \multirow{2}{*}{38.04} & \multirow{2}{*}{171.73} & $\times$ & 1.914 & 75.22 & 2.550 & 84.29 & 2.481 & 83.28 \\
        & & & \checkmark & \textbf{1.963} & \textbf{76.18} & \textbf{2.568} & \textbf{84.43} & \textbf{2.497} & \textbf{83.42} \\    
        \bottomrule
      \end{tabular}
    }
    
    \vspace{1ex}
    {\raggedright \footnotesize
    $^*$ In the D-SKD column, ``Teacher'' refers to the single-channel model. ``$\times$'' and ``\checkmark'' denote the dual-channel student model trained without and with the D-SKD framework, respectively.\par}
    
  \end{minipage}
\end{table}

\section{Conclusion}

In this paper, we propose the D-SKD framework to address the challenge of spatial ambiguity in lightweight dual-channel speech enhancement. By introducing a dynamic arbitration mechanism into the distillation process with a spatially invariant single-channel teacher, D-SKD enables the dual-channel model to adaptively regulate reliance on spatial cues without explicit angular supervision. Experiments show that D-SKD effectively enhances the performance of DC-GTCRN and other SOTA lightweight models in closely-spaced source scenarios while largely preserving performance in other conditions. Notably, our framework requires no additional inference overhead or auxiliary inputs, making it suitable for lightweight deployment on edge devices.

\section{Acknowledgments}

This work is supported by the National Natural Science Foundation of China (Grant No.~12274221) and the AI \& AI for Science Project of Nanjing University.

\section{Use of Generative AI Disclosure}
The authors state that no generative AI tools were used to generate the primary content of this manuscript. Generative AI tools were exclusively employed for language polishing, academic editing, and \LaTeX\ formatting. No novel hypotheses or data generation were involved in this process. All authors assume full responsibility for the content and originality of the paper.

\bibliographystyle{IEEEtran}
\bibliography{references}

\end{document}